\documentclass{acm_proc_article-sp}
\usepackage{graphicx} 
\usepackage{caption}
\usepackage{subcaption}
\usepackage{multicol}
\usepackage[procnames]{listings}
\usepackage{tabularx}
\usepackage{color}
\usepackage{listings}  
\usepackage{tabularx}
 \DeclareCaptionType{copyrightbox}
\definecolor{mygreen}{rgb}{0,0.6,0}
\definecolor{gray}{rgb}{0.4,0.4,0.4}
\definecolor{darkblue}{rgb}{0.0,0.0,0.6}
\definecolor{purple}{rgb}{0.2,0.1,0.3}
\definecolor{background}{rgb}{0.95,0.95,0.95}
\definecolor{myblue}{RGB}{20,105,176}
\lstdefinestyle{styOMP2MPI}{
  language=C++,
  basicstyle=\tiny,
  backgroundcolor=\color{background},
  numbers=left,
  directivestyle=\textbf,
  keywordstyle=[1]\color{mygreen},
  keywordstyle=[2]\color{purple}\textbf,
  keywordstyle=[3]\color{darkblue}\textbf,
  keywordstyle=[4]\color{mygreen}\textbf,
  identifierstyle=\color{black},
  commentstyle=\color{gray},
  stringstyle=\color{myblue},
  moredirectives={omp, hmpp,  target, CUDA, synchronize, delegatedstore, advancedload, parallel,for, codelet, callsite, private, shared, args, addr,  asynchronous, hmppcg, gridify, release, asynchronous, reduction}, 
  keywords=[1]{offset,partSize, MPI_Recv, MPI_Send, followIN, work, killed, stat},
  morekeywords=[2]{in, inout,io}, 
  morekeywords=[3]{group0_46, group, mapbyname, noupdate, true}, 
  morekeywords=[4]{diffsum , diffsum_reduced, reduce}
  deletekeywords={for,int,double, return, true},
  emph={mpi},
  emphstyle=\color{darkblue}\textbf,
  literate=
            *{\{}{{{\color{myblue}{\{}}}}{1}
            {\}}{{{\color{myblue}{\}}}}}{1}
            {[}{{{\color{myblue}{[}}}}{1}
            {]}{{{\color{myblue}{]}}}}{1},
   escapeinside=!!
}

\lstdefinelanguage{omp2mpi}
{
 style=styOMP2MPI,
}

\usepackage[ruled]{algorithm2e}
\SetAlFnt{\algofont}
\SetAlCapFnt{\algofont}
\SetAlCapNameFnt{\algofont}
\SetAlCapHSkip{0pt}
\IncMargin{-\parindent}
\renewcommand{\algorithmcfname}{ALGORITHM}
\newcommand{\includecode}[2][c]{\lstinputlisting[escapechar=]{#2}}

\begin{document}

\title{OMP2MPI: Automatic MPI code generation from OpenMP programs}

%
%
%
%
%

\numberofauthors{3} 
%
\author{
%
%
\alignauthor
Albert Sa\`{a}-Garriga\\
 \affaddr{Universitat Auton\`{o}ma de Barcelona}\\
 \affaddr{Edifici Q,Campus de la UAB}\\
 \affaddr{Bellaterra, Spain}\\
 \email{albert.saa@uab.cat}
\alignauthor David Castells-Rufas\\
 \affaddr{Universitat Auton\`{o}ma de Barcelona}\\
 \affaddr{Edifici Q,Campus de la UAB}\\
 \affaddr{Bellaterra, Spain}\\
 \email{david.castells@uab.cat}
\alignauthor Jordi Carrabina\\
 \affaddr{Universitat Auton\`{o}ma de Barcelona}\\
 \affaddr{Edifici Q,Campus de la UAB}\\
 \affaddr{Bellaterra, Spain}\\
 \email{jordi.carrabina@uab.cat}
\and
}
\date{10 December 2014}

\maketitle
\begin{abstract}
In this paper, we present OMP2MPI a tool that generates automatically MPI source code from OpenMP. With this transformation the original program can be adapted to be able to exploit a larger number of processors by surpassing the limits of the node level on large HPC clusters. The transformation can also be useful to adapt the source code to execute in distributed memory many-cores with message passing support. In addition, the resulting MPI code can be used as an starting point that still can be further optimized by software engineers. The transformation process is focused on detecting OpenMP parallel loops and distributing them in a master/worker pattern. A set of micro-benchmarks have been used to verify the correctness of the the transformation and to measure the resulting performance. Surprisingly not only the automatically generated code is correct by construction, but also it often performs faster even when executed with MPI.

\end{abstract}



\section{Introduction}
One of the strengths of the OpenMP paradigm is the simplicity of its programming model. In it, the invocation of communication primitives are hidden from the programmer as they are implicitly introduced by compilation directives working in conjunction with the OpenMP run time. However, its use is usually limited to shared memory systems. Large HPC systems (like ones in top500 list) are often created by replicating nodes that contain some memory and a number of sockets with multicore processors or accelerators that can access the memory on the node. Memory on remote nodes is not usually visible in the address space on applications running in one node. This makes OpenMP limited to the node domain, making OpenMP applications difficult to scale to a larger number of nodes (and cores) without introducing other paradigms like MPI.

There are runtimes that can overcome this limitation usually by implementing Software Distributed Shared Memory, but they are also transparent to the programmer and, consequently, do not allow any fine tuning that could be needed to better adapt to the potential different contexts. Moreover, they cannot be generally applicable to all distributed memory platforms.

On contrast, MPI is a de facto standard commonly used for big HPC applications. In it, the communication primitives must be explicitly coded. Introducing the communication primitives to implement the cooperation patterns makes the code larger and more difficult to read and understand. Obviously, it is more complex to learn since there are a large number of functions including point to point communication primitives as well as collective communication primitives. This coding effort is justified if it is needed to execute on thousands of cores. MPI allows to communicate among cores on different nodes, and one could think that it introduces performance overheads at the node level compared with OpenMP. But this is a controversial issue with no clear answer as shown in \cite{r1,r2}.

We advocate for a different approach that would let programmers use OpenMP to express the parallelism in their application while automatically generating a MPI equivalent program that can be executed in a distributed memory (DM) machine. A new tool (OMP2MPI) has been developed  which transforms OpenMP source code into MPI source code. The resulting code is valid by construction, and can be executed in different kinds of DM systems, like large HPC clusters or distributed memory experimental processors like Intel Polaris, Ambric, or experimental FPGA based multi-soft-cores (like \cite{r3}). Another potential use is to test if there is any performance gain by using MPI on an application on the same shared memory platform.

The paper is organized as follows, in Section \ref{sec:related} there is a review of the related work, in Section \ref{sec:compiler} compiler transformations done to translate from OpenMP to MPI, the following section present the performance obtained by several automatically created MPI codes from the Polybench benchmark~\cite{poly}. and finally, in Section \ref{sec:conclusion} concludes with an explanation of the obtained results and future tool improvements.

\section{Related Work}
\label{sec:related}
Many source-to-source compiler alternatives have been proposed to the MPI programming complexity, the standard idea is to reuse codes implemented in OpenMP to generate solutions that can be executed using distribute memory architectures. 

Most of the existing projects dedicated to the use of OpenMP codes for distributed memory architectures rely on the use of the software layer to manage data placements on nodes (Software Distributed Shared Memory Architectures). An example of these is OMNI OpenMP\cite{sskt99} and his optimization proposed in \cite{bme02, mms00}, are one way to support OpenMP in a distributed memory environment using a software distributed shared memory system (SDSM) as an underlying run-time system for OpenMP. Cluster-enabled OMNI OpenMP on SCASH is an implementation of OMNI OpenMP compiler for a software distributed shared memory system SCASH running under SCore Cluster System Software.  Another important software system to mark is Cluster OpenMP  proposed by Intel\cite{Hoe06}, that one, as in the aforementioned, allow the use of OpenMP programs to run in clusters, even that was discontinued few years ago. All these solutions, based on software layer, can be used on distributed architectures, without use Message Passing Interface but need some kind of runtime. In contradistinction, OMP2MPI shows the generated solution that will be executed on cluster to the programmer, an this could be optimized, if needed, by an expert offering more flexibility on how will be the code executed in cluster.

More similar ways to port OpenMP programs to Clusters are proposed in PaRADE\cite{kkh03} or based on OMNI compiler, \cite{ehk}, based in Polaris. Both combines the software layer management of data with the use of MPI primitives.

In \cite{DPA08,Gas08,PPJ}, authors propose to extend OpenMP with additional clauses necessary for streamization as in our tool. Nevertheless, the most similar tools are proposed in\cite{be05,bme02} and \cite{step}. Both, are source-to-source compilers as our tool, the first based on Cetus\cite{cetus} and the second on PIPS\cite{pips} generating solutions that could be compared to ours. 

OMP2MPI is based on Mercurium Framework since it supports C/C++ source codes and gives an intermediate representation more friendly to work than the other existing frameworks as LLVM \cite{llvm}, PIPS , Cetus or ROSE \cite{rose}. And have a well documented API that allows to extend that one. 

\section{OMP2MPI Compiler}
\label{sec:compiler}
OMP2MPI is a Source to Source compiler (S2S) based on BSCs Mercurium framework~\cite{Mercurium} that generates MPI code from OpenMP. Mercurium~\cite{balart2004nanos} gives us a source-to-source compilation infrastructure aimed at fast prototyping and supports C and C++ languages. This platform is mainly used in the Nanos environment to implement OpenMP but since it is quite extensible it has been used to implement other programming models or compiler transformations as has been demonstrated in\cite{asaa}, providing OMP2MPI with an abstract representation of the input source code: the Abstract Syntax Tree(AST). AST provides an easy access to source code structure representation, the table of symbols and the context of these.

The specialization of Mercurium for OMP2MPI compiler is achieved using a plugin architecture, where plugins represent several phases of the compiler. These plugins are written in C++ and dynamically loaded by the compiler according to the selected configuration. Code transformations are implemented to the source code which implies that there is no need to know or modify the internal syntactic representation of the compiler. 

Figure \ref{fig:compiler} shows a simplified process flow of OMP2MPI compiler, where an OpenMP input code is transformed in an MPI code by the use and analysis of the AST. OMP2MPI detect and transform OpenMP blocks (focused in \textit{\#pragma omp parallel for}), dividing the task in MPI master and slave processes that will be distributed on the available cores. To determine the OpenMP blocks that have to be transformed OMP2MPI use the directives proposed in~\cite{ayguade2009proposal}, as is illustrated in the input code example shown in Table~\ref{src:omp}. 

The proposed tool is able to use the combination of peer to peer communication functions (MPI\_Send, MPI\_Recv), and divide the code into sequential and parallel parts with the use of MPI ranks.

\begin{figure}[ht!]
\captionsetup[subfigure]{position=bottom}
\centering
        \begin{subfigure}[b]{0.49\columnwidth}
                \includegraphics[width=\columnwidth]{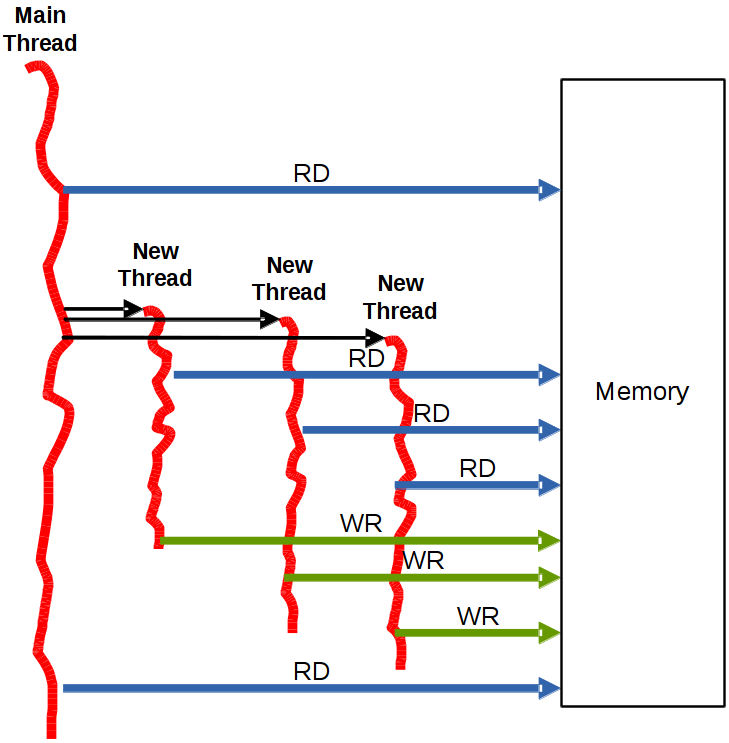}
                \caption{Example of a memory access pattern of an OpenMP application. Threads directly access the shared memory.\\\\\\\\}
                \label{fig:vsomp}
        \end{subfigure}
        \begin{subfigure}[b]{0.49\columnwidth}
                \includegraphics[width=\columnwidth]{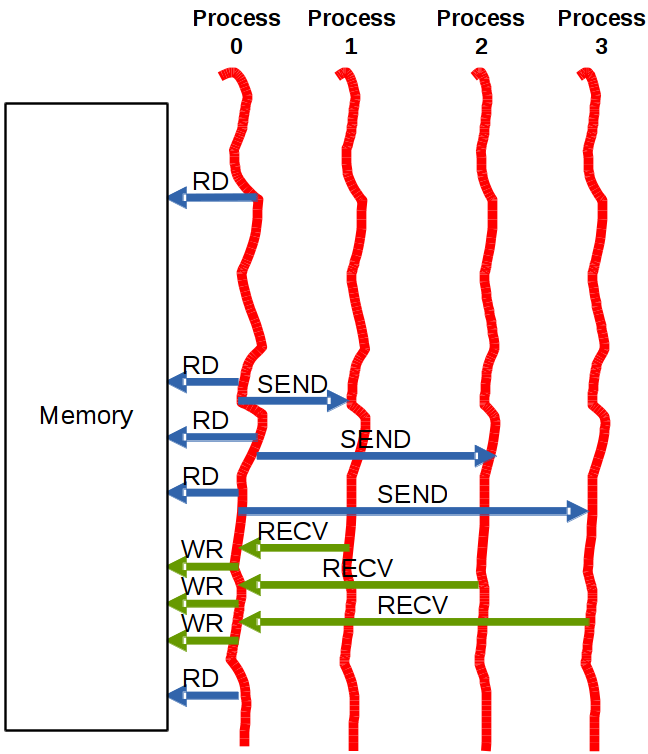}
                \caption{Example of the proposed memory access pattern for shared variables in MPI target applications. Access to shared variables are centralized from Master node, and worker processes have to communicate with it to access them.}
                \label{fig:vsmpi}
        \end{subfigure}
        \caption{Shared memory access on different architectures(Blue lines represent a read operation. Red lines represent a write operation).}
\end{figure}

With these MPI functions OMP2MPI is able to create a correct implementation of a MPI parallel program that in the studied will result similar to an MPI hand-coded version of the original problem. OMP2MPI transforms the original code doing the MPI initialization and workload distribution based on the process rank of the calling process in the communicator. The master process with rank 0 will contain all the sequential code from the original OpenMP application and will manage the shared memory access being the responsible to keep this updated on all the slaves as is shown in Figure \ref{fig:vsmpi}, in contrast to the original OpenMP memory access represented in Figure \ref{fig:vsomp} where all the created threads have access to shared memory. In these figures, lines in blue represent a read operation while lines in read represent a write operation.

\lstset{language=omp2mpi} 
\begin{table}[ht!]
\begin{tabularx}{\columnwidth}{|X|}
  \hline
  \includecode{src/omp.c} \\
  \hline
\end{tabularx}
\caption{OpenMP blocks source code example using the created \textit{target} clause.}
\label{src:omp}
\end{table}

\begin{table*}[!ht]
\begin{tabularx}{\textwidth}{|X|}
  \hline
  \includecode{src/mpi3.c} \\
  \hline
\end{tabularx}
\caption{Resulting piece of code from the transformation of the first OpenMP block shown in Table \ref{src:omp} into MPI Source Code Example that contains the calculation of an Array. In green inserted MPI funcions and created variables.}
\label{src:mpi1}
\end{table*}

\begin{table*}[!ht]
\begin{tabularx}{\textwidth}{|X|}
  \hline
  \includecode{src/mpi2.c} \\
  \hline
\end{tabularx}
\caption{Resulting piece of code from the transformation of the second OpenMP block shown in Table \ref{src:omp} that contains the calculation of a reduced variable into MPI Source Code Example.  In green inserted MPI functions and created variables}
\label{src:mpi2}
\end{table*}

\subsection{AST Manipulation}
\begin{figure}[!ht]
\center
\includegraphics[width=\columnwidth]{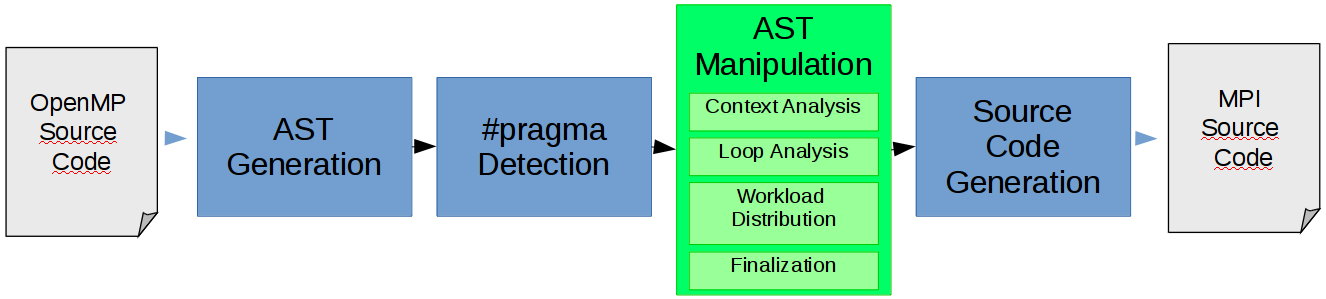}
\caption{In blue, the functionalities already offered by the Mercurium framework. In green the AST manipulation process done by OMP2MPI.}
\label{fig:compiler}
\end{figure}

The AST manipulation stage on Figure \ref{fig:compiler} is composed by four main steps : 1) Context Analysis, 2)Loop Analysis, 3)Workload Distribution, 4)Finalize.

\subsubsection{Context Analysis}
\begin{figure}[!ht]
\center
\includegraphics[width=0.9\columnwidth]{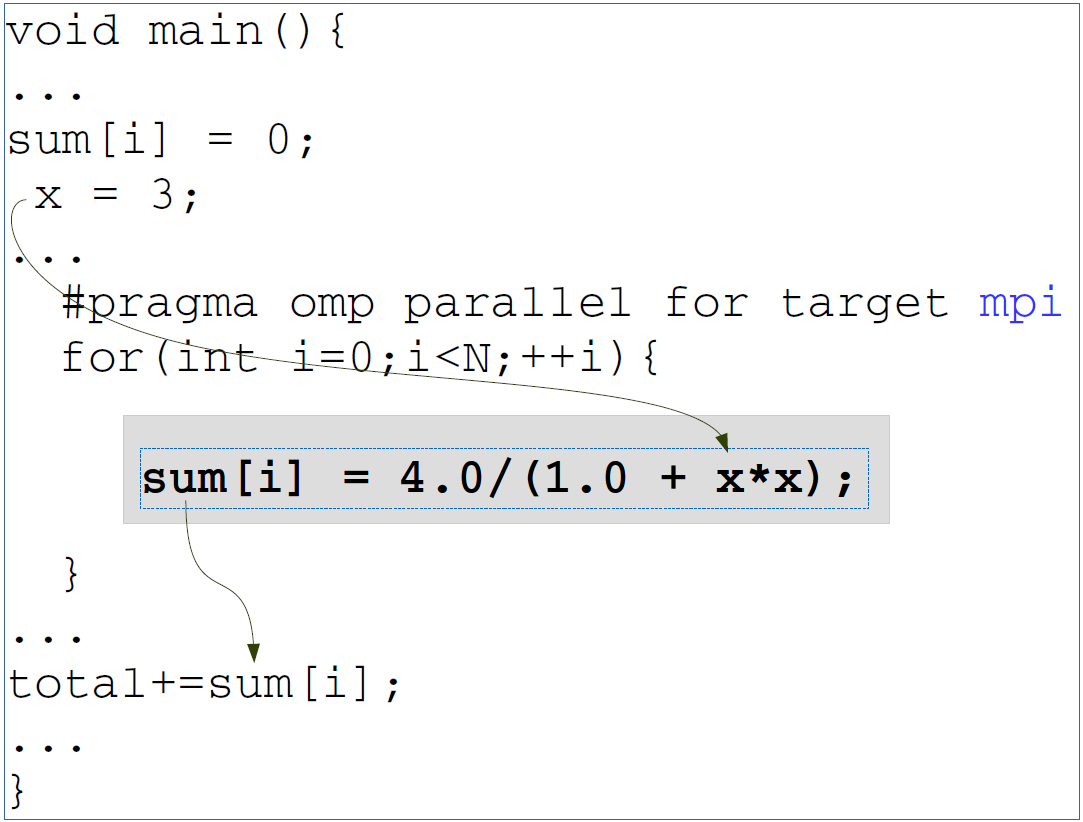}
\caption{Context example. Is possible to see that variable \textit{x} is written before the parallel loop and it is not accessed after it, so it can be labeled as an IN variable. While sum is both written before but not read inside, and read after the parallel loop, so it can be labeled as OUT.}
\label{fig:toy}
\end{figure}

To transform the original code OMP2MPI analyze the context where the OpenMP block is originally computed, and do an accurate contextual analysis of the AST for each of the variables needed inside it. On MPI each of the executed process manage their private variables independently and the main problem to transform OpenMP to MPI is on shared variables, for this reason OMP2MPI study each of shared variables used inside an OpenMP block and analyze the AST to identify when/whether they are accessed. OMP2MPI distinguish the used variables on an OpenMP blocks into IN variables (variables that are read inside the block but without modification), OUT variables (variables that are write inside the block and the result of these are needed after the block finalization) and, INOUT variables (complains both cases). Figure \ref{fig:toy} represents the first OpenMP block implemented in Table \ref{src:omp}, this figure is useful to show the difference between a variable \textit{x} that will be read inside the OpenMP block without any modification(in variable), an \textit{sum} that will be write inside the OpenMP block and will be necessary to have this variable updated before the next read of this(out variable). Depending on that information MPI\_Send / MPI\_Recv instructions are inserted to transfer the data to the appropriate slaves.

The context analysis stage will also include the study of the context situation of the OpenMP block, as could be to detect that the OpenMP block to transform is inside a loop, in which case OMP2MPI will modify where will be inserted the initialization and the task synchronization instructions.

\subsubsection{Loop Analysis}
This stage is the dedicated to study the loop that is included in \textit{pragma omp for} directive to divide correctly the computation of the for loop inner statements. OMP2MPI do an exhaustive analysis of the for semantics understanding and determining which is: 1) The variable iterated, 2) The variable initial value, 3) The variable final value 4) The decrement/increment after each iteration 5) The logic comparison operation. However, there are some cases in which OMP2MPI will not be able to transform loops based on the for loop semanics i.e complex not linear increments on iterator or multiple cases on condition. This cases will do that the studied blocks will not be transformed by OMP2MPI, keeping these as OpenMP blocks.

\subsubsection{Workload Distribution}
\begin{figure}[!ht]
\center
\includegraphics[width=0.75\columnwidth]{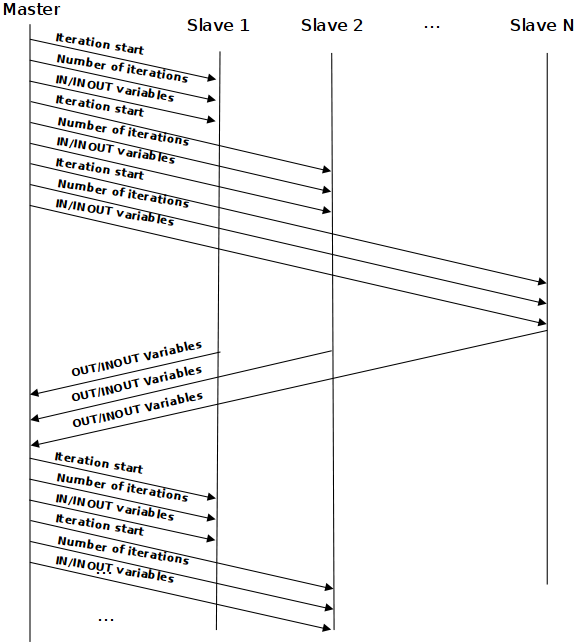}
\caption{Workload static distribution. The work is sent in an orderly manner depending on the rank of slaves. All slaves has to finish before continue with the next piece of workload.}
\label{fig:ioVarsS}
\end{figure}

\begin{figure}[!ht]
\center
\includegraphics[width=0.75\columnwidth]{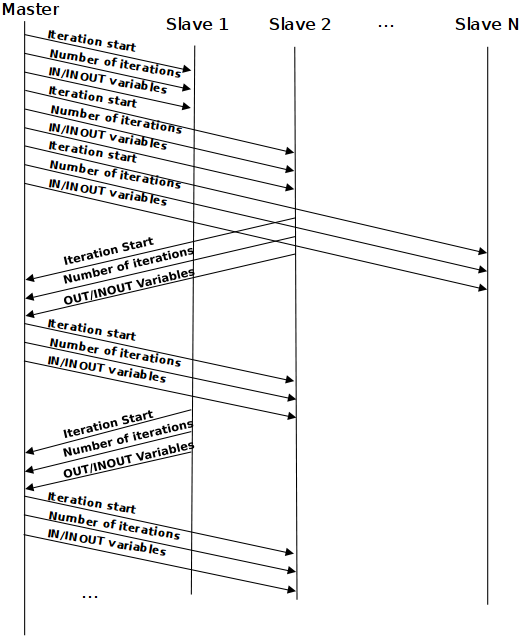}
\caption{Workload dynamic distribution. The work is divided dynamically responding to the slave that answers with the following range of iterations and the variables needed to do the computation.}
\label{fig:ioVarsD}
\end{figure}

Having the context understanding and the proper loop semantics, OMP2MPI will divide the OpenMP block calculation to work with master/slaves MPI model, by using the producer/consumer paradigm. OMP2MPI treated all the variables studied in the context analysis stage . Figures \ref{fig:ioVarsS} and \ref{fig:ioVarsD} shows how OMP2MPI divide the computation for each of the OpenMP block. The iterations of the OpenMP block will be divided in a different way depending if the original OpenMP block contain in his pragma directives a \textit{schedule} clause, as static or guided. OMP2MPI divides the iterations of the for loop between all the available slaves. Figure \ref{fig:ioVarsD} shows the division model for an OpenMP dynamic block  while, \ref{fig:ioVarsS} how an  OpenMP static or guided block will be divided on master/slaves. 

Using the static division the outer loop is scheduled in a round robin fashion by using MPI\_Recv from specific ranks. This could lead to an unbalanced load. However, is necessary to have this kind of division because OMP2MPI is thought to be faithful with the original OpenMP code which could have this directive. On the other hand, in the case of the dynamic division, the outer loop is scheduled dynamically by using ANY\_SOURCE MPI\_Recv and results more efficient. Trying to overcome unbalanced load, OMP2MPI determine the range of iterations that will compute each process on execution by dividing the number of total iterations by the available slaves, and this number is finally divided by 10, as is shown in line 4 of Table \ref{src:mpi1}. Table \ref{src:mpi1} illustrate an example of an array computation, OMP2MPI transforms the first OpenMP block on Table \ref{src:omp} into the showed source code. 
Two different rules to ensure that the calculation over a variable could be divided in independent executions of the original for loop are defined. 
In the case that the divided iterator is linear in the first dimensional access pointer in a write operation of a variable(i.e. \textit{var[i][j]=2*i}), MPI\_Send and MPI\_Recv functions will transfer to the master, just the portion of the out variable that has to be read or has been modified, from offset to the actual maximum iterator. The other studied case is when the divided iterator is not the first dimensional access pointer in a write operation but is used as that in the any of the variables on the assign operation(i.e. \textit{var[i]=2*j}). OMP2MPI will transfer the full array but just in the case that the actual iteration is the last slave in execution.

The used workload distribution is not applicable to all the possible cases that are accepted in an OpenMP block, OMP2MPI is not able to divide the computation on variables with concurrent accesses to a shared variable, when the iterator is on second pointer of in access to that one, or when the variable is not linearly accessed.  

An special case of INOUT variable is the variable that is specified as reduced variable by the OpenMP reduction clause. In this case, OMP2MPI determine the starting value of the reduced variable, depending on the reduction operation(an starting value of 0 for "+" and "-" operations, or 1 for "*" and "/") and will accumulate on the resulting variable the received results computed on slaves by the use of the operation to reduce. Table \ref{src:mpi2} show how that is preformed by OMP2MPI transforming the second block on Table \ref{src:omp} into the showed source code.

\subsubsection{Finalization}
The final stage on the AST Manipulation step is the finalization stage that is responsible to assign the remaining non MPI parallelized source code to the master node to avoid unnecessary computation, and put \textit{MPI\_Finalize} instruction before that. The resulting process is illustrated in the last lines of Table \ref{src:mpi2}.

\section{Results}
\label{sec:res}
\begin{figure}[ht!]
\centering
        \begin{subfigure}[b]{0.45\columnwidth}
                \includegraphics[width=\columnwidth]{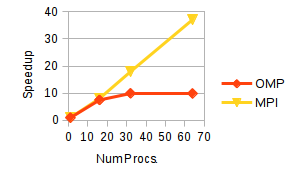}
                \caption{GEMM}
                \label{fig:GEMM}
        \end{subfigure}
        \begin{subfigure}[b]{0.45\columnwidth}
                \includegraphics[width=\columnwidth]{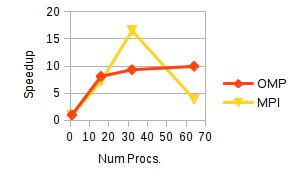}
                \caption{2mm}
                \label{fig:2MM}
        \end{subfigure}
        \begin{subfigure}[b]{0.45\columnwidth}
                \includegraphics[width=\columnwidth]{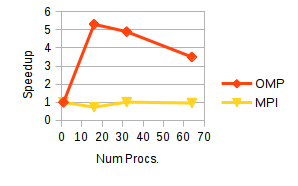}
                \caption{Convolution}
                \label{fig:convolution}
        \end{subfigure}
         \begin{subfigure}[b]{0.45\columnwidth}
                \includegraphics[width=\columnwidth]{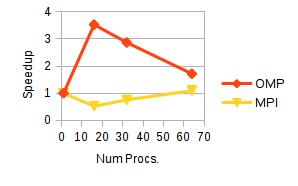}
                \caption{Jacobi 2d}
                \label{fig:jacobi}
        \end{subfigure}
        \begin{subfigure}[b]{0.45\columnwidth}
                \includegraphics[width=\columnwidth]{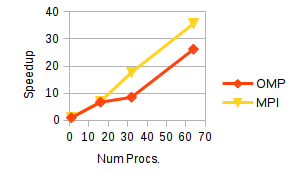}
                \caption{Syrk}
                \label{fig:syrk}
        \end{subfigure}
         \begin{subfigure}[b]{0.45\columnwidth}
                \includegraphics[width=\columnwidth]{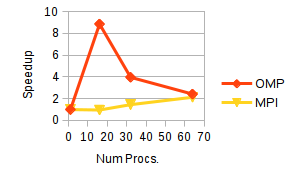}
                \caption{Seidel}
                \label{fig:seidel}
        \end{subfigure}
         \begin{subfigure}[b]{0.45\columnwidth}
                \includegraphics[width=\columnwidth]{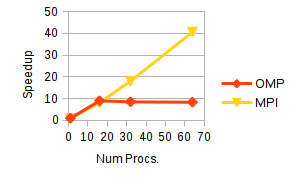}
                \caption{Syr2k}
                \label{fig:syr2k}
        \end{subfigure}
        \caption{Speed-up of the tested problems by using 16, 32 and, 64 processors compared with the sequential version.}
        \label{fig:results}
\end{figure}

We compiled with OMP2MPI a subset of the Polybench benchmark. The generated versions were executed in  64 CPU's  E7-4800 with  2.40 GHz(Bullion quadri module) and compiled with bullxmpi, compatible with MPI 2.1, enhanced by Bull with many new features such as effective abnormal pattern detection, network-aware collective operations, and multi-path network fail-over, to increase reliability, resilience and boost the performance of parallel MPI applications. We compare the codes resulting from the execution of OMP2MPI with the original OpenMP ones, and also with a sequential version of the same problem. Figure~\ref{fig:results} shows the speed-up comparison for the selected problems. This figure shows that OMP2MPI produces good transformation of the original OpenMP code, and in most of cases the generated have better scalability than the original one with linear speed-up increment correlated with the number of processors used on execution.

\section{Conclusions}
\label{sec:conclusion}
We have presented OMP2MPI, a tool that facilitates the portability of an OpenMP source code to MPI, we shown how it effectively automatically translates OMP2MPI being able to go outside the node. Allowing that the program exploits non shared-memory architectures such as cluster, or NoC-based MPSoC. 

This  automatic task is very useful because  the  programmer could keep working with the OpenMP model, being easily readable and  just  compile  over OMP2MPI compiler  to take the advantages of the MPI model offer (speed-up, scalability, etc.). The readability of the code generated is acceptable so that allows further optimization  by  an  expert intending to improve performance results. The  experiments made using  Polyhedral Benchmark are  promising for this effortless version  and produce better scalability than the original  OpenMP code, Speed-up figures for 64 cores in most of cases are higher than 20$\times$ compared to the sequential version, and also higher  than 4$\times$ compared to the original OpenMP code. These results show again, as mentioned in the introduction, that OpenMP does not always perform better than MPI in shared memory systems.

Future improvements on OMP2MPI will be done to include all the possible uses of shared variables inside OpenMP block and to allow the use of \textit{target mpi} clauses on more OpenMP directives as example on \textit{critical} sections that could be transformed by the use of \textit{MPI\_AllReduce} or \textit{atomic} sections transformed to \textit{MPI\_Bcast}.

\section{Acknowledgments}
This work was partly supported by the European cooperative CATRENE project CA112 HARP, the Spanish Ministerio de Econom\'{i}a y Competitividad project IPT-2012-0847-430000, the Spanish Ministerio de Industria, Turismo y Comercio projects and TSI-020100-2010-1036, TSI-020400-2010-120. The authors thank BULL SAS for their support. 
\bibliographystyle{abbrv}
\bibliography{sigproc}

\end{document}